\documentclass[aps,twocolumn,groupedaddress,showpacs]{revtex4}
\usepackage{amsfonts,amssymb,amscd,amsmath}
\usepackage{graphicx}
\usepackage{epsfig}
\usepackage{latexsym}
\usepackage{amssymb}
\usepackage{subfigure}

\begin{document}

\def\abs#1{ \left| #1 \right| }
\def\lg#1{ | #1 \rangle }
\def\rg#1{ \langle #1 | }
\def\lrg#1#2#3{ \langle #1 | #2 | #3 \rangle }
\def\lr#1#2{ \langle #1 | #2 \rangle }
\def\me#1{ \langle #1 \rangle }

\newcommand{\bra}[1]{\left\langle #1 \right\vert}
\newcommand{\ket}[1]{\left\vert #1 \right\rangle}
\newcommand{\bx}{\begin{matrix}}
\newcommand{\ex}{\end{matrix}}
\newcommand{\be}{\begin{eqnarray}}
\newcommand{\ee}{\end{eqnarray}}
\newcommand{\nn}{\nonumber \\}
\newcommand{\no}{\nonumber}
\newcommand{\de}{\delta}
\newcommand{\lt}{\left\{}
\newcommand{\rt}{\right\}}
\newcommand{\lx}{\left(}
\newcommand{\rx}{\right)}
\newcommand{\lz}{\left[}
\newcommand{\rz}{\right]}
\newcommand{\inx}{\int d^4 x}
\newcommand{\pu}{\partial_{\mu}}
\newcommand{\pv}{\partial_{\nu}}
\newcommand{\au}{A_{\mu}}
\newcommand{\av}{A_{\nu}}
\newcommand{\p}{\partial}
\newcommand{\ts}{\times}
\newcommand{\ld}{\lambda}
\newcommand{\al}{\alpha}
\newcommand{\bt}{\beta}
\newcommand{\ga}{\gamma}
\newcommand{\si}{\sigma}
\newcommand{\ep}{\varepsilon}
\newcommand{\vp}{\varphi}
\newcommand{\zt}{\mathrm}
\newcommand{\bb}{\mathbf}
\newcommand{\dg}{\dagger}
\newcommand{\og}{\omega}
\newcommand{\Ld}{\Lambda}
\newcommand{\m}{\mathcal}
\newcommand{\dm}{{(k)}}

\title{Variational Limits for Phase Precision in Linear Quantum Optical Metrology}
\author{Yang Gao}
\email{gaoyangchang@outlook.com} \author{Ru-min Wang}
\affiliation{Department of Physics, Xinyang  Normal University,
Xinyang, Henan 464000, People's Republic of China}

\begin{abstract}
We apply the variational method to obtain the universal and
analytical lower bounds for parameter precision in some noisy
systems. We first derive a lower bound for phase precision in lossy
optical interferometry at non-zero temperature. Then we consider the
effect of both amplitude damping and phase diffusion on phase-shift
precision. At last, we extend the constant phase estimation to the
case of continuous fluctuating phase estimation, and find that due
to photon losses the corresponding mean square error transits from
the stochastic Heisenberg limit to the stochastic standard quantum
limit as the total photon flux increases.
\end{abstract}

\maketitle

\section{Introduction}

A main task of quantum metrology is to find the limit of precision
in the estimation of parameter $x$ \cite{qq,uncert,HL}. According to
the general quantum estimation theory, a typical parameter
estimation consists in sending a probe in a suitable initial state
through some phase-sensitive physical device and measuring the final
state of the probe. Let $\xi$ be the outcome of the measurement, and
$X(\xi)$ be the estimator of $\xi$ constructed from the outcome $x$.
A local parameter precision of the estimation is quantified by the
uncertainty $\delta x^2 = \int (X(\xi)-x)^2 p(\xi|x)d\xi $, where
$p(\xi|x)$ is the conditional probability distribution of obtaining
a certain outcome $\xi$ given $x$. Better precision is obtained upon
decreasing $\delta x$. The minimization of $\delta x$ over all
possible measurement procedures leads to the quantum Cramer-Rao
inequality $\delta x \ge 1/\sqrt{\nu \m F_Q}$ \cite{qq,uncert}. Here
$\nu$ is the number of repeated measurements and $\m F_Q$ is called
the quantum Fisher information (QFI).

For quantum optical metrology with separable input states, the QFI
for estimating phase $\vp$ gives the standard quantum limit (SQL)
$\delta \vp \sim 1/\sqrt{N}$, where $N$ is the number of resources
utilized in optical interferometer. In the absence of noise,
employing quantum resources such as coherence and entanglement in
the input state, it is possible to hit the Heisenberg limit (HL)
$\delta \vp \sim 1/N$, namely the ultimate phase estimation limit
\cite{HL}. In the presence of noises, it has been shown that the
transition of phase precision from the HL to the SQL can occur with
an increasing $N$ \cite{numer}. However, for noisy systems, most of
known expressions for QFI involve quite cumbersome optimization
procedures when the number of resources increases. In Refs.
\cite{var1,var2}, a general variational method is proposed to obtain
a useful and analytical bound for QFI.

This variational method has been applied to phase estimation with
lossy optical interferometry at zero temperature $T=0$, frequency
estimation with atomic spectroscopy in the presence dephasing
\cite{var1}, phase-shift estimation under phase diffusion
\cite{var2}, and weak classical force estimation \cite{pure}. Such
obtained lower bounds could capture the main features of the
corresponding numerically rigorous results, and thus provide much
useful information for the ultimate limit for parameter precision.

In this paper, we apply the variational method to more noisy systems
and obtain some lower bounds for parameter precision. In Section II,
we first review the variational method proposed in Ref.
\cite{var1,var2}. Then in Section III, a lower bound for phase
precision in lossy optical interferometry at non-zero temperature is
derived through the variational method. Next, in Section IV we
consider the effect of both amplitude damping and phase diffusion on
phase-shift precision. In Section V, the constant phase estimation
is extended to the case of continuous fluctuating phase estimation
\cite{shl}, and find that the mean square error (MSE) that
quantifies the estimation precision transits from the stochastic HL
to the stochastic SQL as the total photon flux increases. Finally,
we end with a short conclusion.

\section{variational method for QFI}
The QFI plays a key role in quantum metrology, but the analytical
expression or numerical calculation of the QFI usually pose
formidable challenges, except for very specific examples when the
density state can be simply put in the diagonal form \cite{exact}.
Recently, an alternative to find an upper bound for QFI was
presented via a variational method in Refs. \cite{var1,var2}. This
method is based on purification technique to optimize the QFI over
all possible purifications of the original input state \cite{inf}.

For a closed system with a pure input state $\rho_0=\lg \psi \rg
\psi$ and a unitary evolution operator $U(x)$ depending on the
unknown parameter $x$, the corresponding QFI can be expressed as $\m
F_Q=4 \Delta H^2$, where $\Delta H^2 = \lrg {\psi} {H^2(x)} \psi
-\lrg {\psi} {H(x)} \psi^2$ with $H(x) = i U^\dg(x) dU(x)/dx $.

For more general situations where the input is a mixture or the
evolution is not governed by a unitary operator, an analytical
expression for $\m F_Q$ remains an open problem. Therefore, an
universal upper bound to $\m F_Q$ based on the convexity of $\m F_Q$
becomes necessary and can always be established for general quantum
channels. If the output state $\rho_S(x)$ of the system is mixed, it
is always possible to enlarge the size of the original Hilbert space
$S$ with an auxiliary environment space $E$ and build a pure state
$\lg {\Phi_{SE}(x)} $ in the enlarged space $S+E$ that satisfies
$\mathrm{Tr}_E \rho_{SE}(x) =\rho_S(x) $, where $\rho_{SE}(x)=\lg {
\Phi_{SE}(x)} \rg {\Phi_{SE}(x)}$. The state $\lg {\Phi_{SE}(x)} $
is called a purification of $\rho_S(x)$.

An upper bound $C_Q$ of $\m F_Q[\rho_S(x)]$ can be obtained: \be C_Q
= \m F_Q[\rho_{SE}(x)] \ge \m F_Q[\rho_{S}(x)]. \ee The physical
reason is that when a system plus an environment are monitored
together, the acquired information on the unknown parameter can not
be smaller than the information obtained when only the system is
measured. It was shown in Ref. \cite{var1} that the QFI can be
attainable: \be \m F_Q[\rho_S(x)]=\mathop\zt {min}_{\lg
{\Phi_{SE}(x)}} C_Q[\rho_{SE}(x)]. \ee Because there is always a
unitary operator $u_E(x)$, acting only on the $E$ space, that
connects two purifications $\lg {\Psi_{SE}(x)}$ and $\lg
{\Phi_{SE}(x)}$ of the same state $\rho_S(x)$, the QFI can be found
by minimizing $C_Q[u_E(x)\rho_{SE}(x)u^\dg_E(x)]$ over all unitary
operators $u_E(x)$ on $E$ space. The effect of $u_E(x)$ is to erase
all non-redundant information on $x$ that has been leaked from space
$S$ into the enlarged space $S+E$.

The equation for $u_E(x)$ optimizing $C_Q$ can be found by
variational method, \be h_E^\zt {opt}\rho_E + \rho_E h_E^\zt {opt} =
\zt{Tr}_S \lz H_{SE} \rho_{SE}+\rho_{SE}H_{SE} \rz, \label{opt} \ee
where $\rho_E(x)=\zt {Tr}_S[\rho_{SE}(x)]$ is the reduced density
matrix in the space $E$. Here two Hermitian operators $h_E$ and
$H_{SE}$ are defined by \be h_E = i {du^\dg_E(x) \over dx} u_E(x),
\nn i{d\lg {\Phi_{SE}(x)} \over dx} = H_{SE}(x) \lg {\Phi_{SE}(x)}.
\ee After determining $u_E^\zt {opt}$ from Eq. (\ref{opt}), the QFI
can be expressed as \be \m F_Q[\rho_S(x)] = C_Q[\rho_{SE}(x)]-4
[\Delta h_E^\zt{opt}(x)]_\Phi^2. \ee Since this method is
variational, whenever it is too difficult to solve Eq. (\ref{opt}),
we can still get some non-trivial and analytical upper bounds to the
QFI. The utility of this type of bounds has been exemplified with
optical interferometry at $T=0$ under photon losses or phase
diffusion, and atomic spectroscopy under dephasing in Refs.
\cite{var1,var2}.

\section{precision limit for lossy interferometry at $T > 0$}

The effect of photon losses in optical interferometry at zero
temperature has been thoroughly investigated in literature, but the
study of the influence of temperature on phase precision is still
rare. Now we apply the variational method to find an analytical
lower bound for phase precision in the presence of photon losses at
$T>0$.

Let the pure input state of mode $a$ be $\lg \psi$, and let
$\rho_S(\vp)$ be the mixed output state after the $\vp$-dependent
dynamical evolution. Under the photon losses and in contact with a
thermal bath $E_1$ of mode $b$ at $T>0$, the dynamical map can
described by the transformation: \be a \to e^{-i \vp} (a \sqrt{\eta}
+b\sqrt{1-\eta}), \ee where $\eta$ quantifies the photon losses from
the lossless case ($\eta=1$) to the complete absorption ($\eta=0$),
and the average excitation number of the thermal bath is assumed as
$\me {b^\dg b} = n_T$. To get an overall purified state, we
introduce a second bath $E_2$ of mode $c$ and take $E_1$ as reduced
from $E_1+E_2$ by the map: $b \to b\sqrt{n_T+1}+c^\dg \sqrt{n_T}$,
where both modes $b$ and $c$ are initially in the ground state. A
possible purification of $\rho_S(\vp)$ can thus be built with an
environment with two baths $E_1$ and $E_2$. This particular
purification consists of three unitary operations on the input state
of $S+E_1+E_2$, \be \lg {\Phi_{SE}(\vp)} = U(\vp) U_1 U_2 \lg
{\Phi_0}, \ee where $U(\vp)=e^{-i\vp a^\dg a}$,
$U_1=e^{\theta_1(ab^\dg-a^\dg b)}$, $U_2=e^{\theta_2(b^\dg c^\dg-b
c)}$, and $\lg {\Phi_0}=\lg \psi_S \lg 0_{E_1} \lg 0_{E_2}$
associated with $\cos\theta_1 =\sqrt{\eta}$ and
$\cosh\theta_2=\sqrt{n_T+1}$.

Any two purifications of a density matrix can be related by a
unitary operator $u_E(\vp)$ acting only on the environment, so we
will adopt this purification as $\lg {\Phi_{SE}(\vp)}$. The most
general purification then takes the form: \be \lg {\Psi_{SE}(\vp)} =
u_E(\vp) \lg{\Phi_{SE}(\vp)}. \ee To make the upper bound $C_Q$ for
the QFI of the system $S$ as lower as possible, we should choose an
proper $u_E(\vp)$ in order to minimize the QFI of $S+E_1+E_2$. Here
$u_E(x)$ is used to erase all non-redundant information on phase
$\vp$ that has been leaked into the environment. Noting the symmetry
between $b$ and $c$, we take a trial form of $u_E$ to counteract the
phase shift on the environment modes, \be u_E(\vp)=e^{-i \vp [\al
b^\dg b+\bt c^\dg c+\ga (bc+b^\dg c^\dg)]}, \ee where $\al$, $\bt$
and $\gamma$ are adjustable variables to minimize $C_Q$. So this
class of purifications of $\rho_S(\vp)$ is given by \be \lg
{\Psi_{SE}(\vp)} = \m U_{SE}(\vp) \lg{\Phi_0}, \ee where $\m
U_{SE}(\vp)=u_E(\vp)U(\vp)U_1 U_2$.

\begin{figure}[t!]
\centering { \label{Fig.1}
\includegraphics[width=.8\columnwidth]{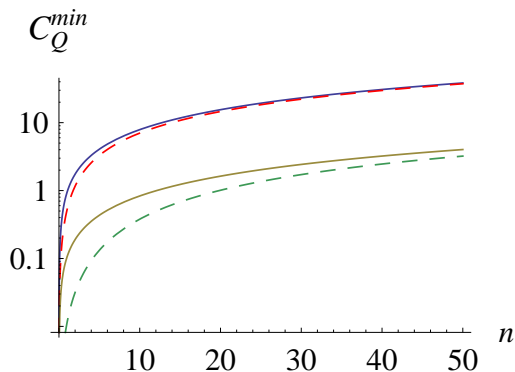}}\hspace{0.00in}
\caption{Comparison between the upper bound $C_Q^\zt{min}$ (solid
lines) and the exact QFI (dashed lines) as a function of the mean
energy $\me n$ for a squeezed vacuum. Here the loss parameter is
$\eta=0.8$. From bottom to top, $n_T=100$ and $n_T=10$.}
\end{figure}

The expression for $C_Q$ can be calculated from \be C_Q=\m F_Q[\lg
{\Psi_{SE}(\vp)}] = 4 \lz \Delta \m H_{SE}^2 \rz_{\Phi_0}, \ee where
\be \m H_{SE}(\vp) = i \m U_{SE}^\dg(\vp)\frac{d \m
U_{SE}(\vp)}{d\vp} \ee is the effective Hamiltonian in the enlarged
space $S+E_1+E_2$. Using the definition of $\m U_{SE}(\vp)$, the
explicit form of $\m H_{SE}$ is found as \be \m H_{SE} = U_2^\dg
U_1^\dg [a^\dg a+\al b^\dg b+\bt c^\dg c \nn +\ga (bc+b^\dg
c^\dg)]U_1 U_2, \ee which can be further simplified by the relations
\be U_1^\dg \lx
\bx a \\ b \ex \rx U_1 &=& \lx \bx \cos \theta_1 & \sin\theta_1 \\
-\sin \theta_1 & \cos\theta_1 \ex \rx \lx \bx a \\ b \ex \rx, \nn
U_2^\dg \lx \bx b \\ c^\dg \ex \rx U_2 &=& \lx \bx \cosh \theta_2 &
\sinh\theta_2 \\ \sinh \theta_2 & \cosh\theta_2 \ex \rx \lx \bx b
\\ c^\dg \ex \rx.
\label{ident} \ee

The final result of $C_Q$ is given by \be C_Q=\Delta n^2 \left(c_1^2
+\al s_1^2 \right)^2 +{\me n} s_1^2[c_1c_2(1-\al)-\gamma s_2]^2 \nn
+ (\me n+1) s_1^2[c_1 s_2(1-\al)-\ga c_2]^2 \nn+ [(s_1^2+\al
c_1^2+\bt)c_2s_2+\ga c_1(c_2^2+s_2^2)]^2, \ee where $\Delta n^2=\lrg
\psi {n^2} \psi -\me n^2$, $\me n =\lrg \psi n \psi$ for the number
operator $n=a^\dg a$, and the conventions $c_1=\cos \theta_1$,
$s_1=\sin \theta_1$, $c_2=\cosh \theta_2$, and $s_2=\sinh \theta_2$
are used. The minimization of $C_Q$ over $\al$, $\bt$ and $\ga$
leads to \be C_Q^\zt {min} = \frac{4}{{1 \over \Delta n^2}+{1-\eta
\over \eta}\lx {n_T+1 \over \me n}+{n_T \over \me n+1}\rx}.
\label{tem} \ee This universal analytical upper bound to the QFI of
the system allows us to evaluate the effect of temperature in the
lossy interferometry irrespective of the input state. At zero
temperature ($n_T=0$), it becomes \be C_Q^\zt {min} = \frac{4}{{1
\over \Delta n^2}+{(1-\eta) \over\eta \me n}}, \ee which is the same
as the result in Ref. \cite{var1}. When the temperature goes to
infinity ($n_T \to \infty$), Eq. (\ref{tem}) implies $\delta \vp \to
\infty$, i.e., no phase information can be obtained.

For the input squeezed vacuum \cite{book}, $\lg \psi=e^{r(a^{\dg
2}-a^2)/2}$ with $\me n=\sinh^2 r$ and $\Delta n^2=2\me n(\me n+1)$,
the exact QFI can be obtained by \cite{me} \be \m
F_Q[\rho_S(\vp)]=\frac{4 u^2}{1+v^2-u^2}, \ee where $u=\eta \sinh
2r$ and $v=\eta \cosh2r+(1-\eta)(2n_T+1)$. The comparison of
$C_Q^\zt{min}$ with this exact QFI is shown in Fig. 1. It can be
seen that the higher the temperature is, the worse the phase
precision is. We can also see that the variational bound saturates
the exact QFI as $\me n \to \infty$.

\section{precision limit under damping and diffusion}

Besides photon losses, phase diffusion is another important source
of noise in optical phase estimation and should be taken into
account. A numerical study of the influence of phase diffusion on
the optimal phase precision was presented in Ref. \cite{dephase}.
Then an analytical lower bound was found by the above variational
method in Ref. \cite{var2}. However, the effect of photon losses is
not considered at the same time. In this section, we will discuss
the effect of both photon losses and phase diffusion on the limit of
phase precision.

For simplicity, suppose a pure input state $\lg \psi$ of mode $a$,
which undergoes a phase shift $\vp$ due to some physical process. In
the presence of photon losses and phase diffusion, a specific
purification of the output state $\rho_S(\vp)$ is \be \lg
{\Phi_{SE}(\vp)} = U(\vp) U_\theta U_\ld \lg {\Phi_0}, \ee where
$U_\theta=e^{\theta(ab^\dg-a^\dg b)}$ and $U_\ld= e^{i (2\ld) a^\dg
a x_2}$ with $\cos \theta = \sqrt{\eta}$, $x_2=(c+c^\dg)/\sqrt{2}$,
and $\ld$ characterizes the strength of phase diffusion in the
process \be \rho_S \to \frac{1}{\sqrt{4\pi \ld^2}}\int e^{-{\phi^2
\over 4\ld^2}} U^\dg(\phi)\rho_S U(\phi)d\phi. \ee In terms of the
Fock basis of the system, the action of phase diffusion takes \be
\lg l\rg k \to e^{-\ld^2 (k-l)^2}\lg l\rg k. \ee

In order to get a tighter upper bound to $\m F_Q[\rho_S(\vp)]$, we
take a trial form of $u_E(\vp)$ inspired by Ref. \cite{var2}, \be
u_E(\vp)=e^{-i \al \vp b^\dg b+i \bt \vp p_2/(2\ld)}. \ee This
choice of $u_E(\vp)$ leads to $C_Q=4[\Delta \m H_{SE}^2]_{\Phi_0}$,
where $ \m H_{SE} = U^\dg_\ld U^\dg_\theta [a^\dg a + \al b^\dg
b-\bt p_2/(2\ld)] U_\theta U_\ld$ with $p_2=(c-c^\dg)/(i\sqrt{2})$.
Using the identity (\ref{ident}) and $U^\dg_\ld p_2 U_\ld = p_2 + 2
\ld a^\dg a$, we have \be C_Q=4\Delta n^2[\eta+\al (1-\eta)-\bt]^2
\nn +4\me n (1-\al)^2 \eta (1-\eta)+\frac{\bt^2}{2\ld^2}. \ee The
minimization of $C_Q$ over the variables $\al$ and $\bt$ gives \be
C_Q^\zt {min} = \frac{4}{\frac{1}{\Delta n^2}+{(1-\eta) \over\eta
\me n}+8\ld^2}. \ee For $\eta=1$, it gives $\de \vp^2 \ge 1/(4\Delta
n^2)+2\ld^2$, which is the same as the result in Ref. \cite{var2}.

Following the steps in Section III, we can also include the effect
of temperature during the process of photon losses. Through some
calculations, we obtain the phase precision \be \de \vp^2 \ge {1
\over 4\Delta n^2}+{1-\eta \over 4\eta}\lx {n_T+1 \over \me n}+{n_T
\over \me n+1}\rx+2\ld^2. \label{boud} \ee The presence of a
constant term in this equation means that phase diffusion is usually
more harmful to phase precision than photon losses.

\begin{figure}[t!]
\centering { \label{Fig.1}
\includegraphics[width=.7\columnwidth]{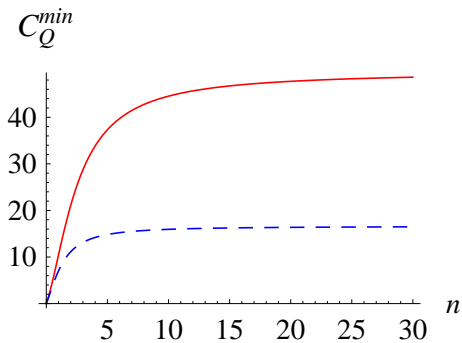}}\hspace{0.00in}
\caption{Comparison between the upper bound $C_Q^\zt{min}$ (solid
lines) and the lower bound $\m I_Q^\zt{opt}$ (dashed lines) as a
function of the mean energy $\me n$ for a squeezed vacuum. Here the
loss parameter and diffusion strength are $\eta=0.95$ and $\ld=0.1$,
respectively. The temperature is set to zero. }
\end{figure}

For the input squeezed vacuum, the exact QFI is not known
analytically. Instead, a lower bound $\m I_M$ of the QFI can be
deduced from the measurement $M=i(a^2-a^{\dg 2})$ \cite{me}. It
gives \be \m I_M=\frac{|\p_\vp \me M_{\rho_S}|^2}{[\Delta
M^2]_{\rho_S}}, \ee where $\me M_{\rho_S} = u e^{-4\ld^2}
\sin(2\vp)$ and \be {[\Delta M^2]_{\rho_S}} &=& 1+v^2+{u^2 \over
2}[1+e^{-8\ld^2} \nn && -e^{-8\ld^2}(1+3e^{-8\ld^2}) \cos(4\vp)].
\ee The optimization of $\m I_M$ over $\vp$ corresponds to $\vp=0$,
and \be \m I_M^\zt{opt}=\frac{4u^2
e^{-8\ld^2}}{1+v^2+{u^2}(1-3e^{-8\ld^2})/2}. \ee The comparison of
$C_Q^\zt {min}$ in Eq. (\ref{boud}) with $\m I_M^\zt{opt}$ is
plotted in Fig. 2. We see that the upper bound $ C_Q^\zt{min}$ to
the QFI is always larger than its lower bound $\m I_M^\zt{opt}$ and
the exact QFI should lay between them. It also displays that both of
bounds approaches to some constant value as the total available
energy go to infinity.

\section{point estimation precision limit for a continuous fluctuating phase}

In the case of estimating a constant phase, the fundamental bound to
phase precision is the HL, which is a quadratic improvement over the
SQL. Contrary to this constant situation, the fundamental bound to
estimating Wiener phase fluctuations $\vp(t)$ with a beam $a(t)$
(obeying $[a(t),a^\dg(t')]=\de (t-t')$) is the stochastic HL, which
shows a MSE scaling as $\mathcal N^{-2/3}$, while the stochastic SQL
scales as $\mathcal N^{-1/2}$ \cite{shl}. Here the beam is assumed
as Gaussian stationary statistics, and $\mathcal N = \me
{a^\dg(t)a(t)}$ is the mean flux (photons per second) in the beam.
In this section, we will show a transitive behavior of MSE scaling
from the stochastic HL to the stochastic SQL as the beam undergoing
some photon losses.

In Ref. \cite{defer}, a continuous form of the quantum Cramer-Rao
inequality was derived, giving a lower bound on the MSE of any
unbiased estimator $\hat \vp(t)$, of a time-varying phase $\hat
\vp(t)$, \be \me { [\hat \vp(t)- \vp(t)]^2 } \ge  \mathcal
F^{-1}(t,t). \label{mse} \ee Here $\mathcal F(t,t')$ is the Fisher
information matrix (with continuous indices $t$ and $t'$) of the
phase of the beam. It includes the sum of quantum and classical
parts $\m F(t,t') = \mathcal F_Q(t,t') + \mathcal F_C(t,t')$. The
(matrix) inverse in Eq. (\ref{mse}) is defined by $\int \mathcal
F^{-1}(t,s) \mathcal F(s,t')ds = \delta(t-t')$. Because the beam is
stationary, all quantities dependent on two times $t$ and $t'$ are
only functions of $t-t'$.

In order to determine $\mathcal F^{-1}(0)$, put Eq. (3) into Eq. (4)
and take the Fourier transform, to give be \be \tilde{\m F}^{-1}
(\og) = {1 \over\tilde{ \m F}_Q(\og) +\tilde{ \m F}_C(\og)} \ee for
the Fourier transform of $\m F^{-1}(t - t')$.  The value of
$\mathcal F^{-1}(0)$ is then given by \be \mathcal F^{-1}(0) =
\frac{1}{2\pi}\int {1 \over\tilde{ \m F}_Q(\og) +\tilde{ \m
F}_C(\og)} d\og .\ee In Refs. \cite{shl}, an upper bound to
$\mathcal F_Q(t,t')$ is taken as $4\me {\Delta n(t)\Delta n(t')}$,
corresponding to the lossless case, and the Wiener phase spectrum
$\tilde{\mathcal F}_C (\og) =\kappa^{p-1}/|\og|^p$ ($p>1$) is
assumed. Considering the analytical properties of $\tilde{ \m
F}_Q(\og)$ for the stationary Gaussian beam, it was shown that the
MSE scales as $\m N^{2(p-1)/(p+1)}$.

However, when the beam suffers some photon losses, the above lower
bound to the MSE is not tight. Similar to the case of constant
phase, we introduce an vacuum environment $b(t)$ (obeying
$[b(t),b^\dg(t')]=\delta (t-t')$ and $\me {b^\dg(t)b(t)}=0$) to
purify the mixed output state. This vacuum is coupled with the beam
through the unitary operator $U_\theta = \exp\lz\theta \int
[a(t)b^\dg(t)-a^\dg(t)b(t)] dt\rz $ with $\cos \theta =
\sqrt{\eta}$. To find a upper bound $C_Q(t,t')$ of $\m F_Q(t,t')$,
we apply a variational unitary operator $\exp[ -i\bt \int \vp(t)
b^\dg(t)b(t) dt]$ on the purified state with $\bt$ as a variational
parameter. The total transformation is thus given by \be \m U_{SE}=
\exp \lz -i \int \vp(t) [a^\dg(t)a(t)+\bt b^\dg(t)b(t)] dt \rz
U_\theta, \ee and the upper bound to $\m F_Q(t,t')$ follows \be
C_Q(t,t')= 4\me {\Delta n(t)\Delta n(t')}[\eta+\bt(1-\eta)]^2\nn +
4\m N \delta (t-t') (1-\bt)^2\eta (1-\eta). \ee Since $\bt$ is
arbitrary, we set $\bt=\eta/(\eta-1)$ and $C_Q$ becomes \be
C_Q={4\eta\m N \over (1-\eta)} \delta (t-t'). \ee This implies that
\be \m F^{-1}(0) &\ge & {1\over2\pi}\int_{-\infty}^\infty
\frac{\kappa^{p-1}}{|\og|^p+\kappa^{p-1}{4\eta\m N \over (1-\eta)} }
d\og \nn &=& \m O\lx(\m N/\kappa)^{1-p \over p}\rx, \ee which is
just the stochastic SQL. In other words, in presence of photon
losses, the special quantum strategy to estimate a fluctuating phase
does not provide an order of magnitude improvement with respect to
the standard light beams.

To illustrate the transitive behavior of the MSE scaling, we use the
analytical property of the Fourier transform of $\me {\Delta
n(t)\Delta n(t')}$ under the same conditions as in Refs. \cite{shl},
and find that \be \m F^{-1}(0) &\ge& \mathop{\max_{\lt\bt\rt}}
{1\over2\pi}\int_{-\infty}^\infty
\frac{\kappa^{p-1}}{|\og|^p+\kappa^{p-1}\tilde{C}_Q(\og)} d\og \\
&\ge & {1\over \pi}\int_L^\infty \frac{\kappa^{p-1}}{\og^p+D} d\og
\nn &\ge& {1\over \pi}\int_0^\infty \frac{\kappa^{p-1}}{\og^p+D}
d\og-{1\over \pi}\int_0^L \frac{\kappa^{p-1}}{D} d\og \nn &\ge& \m
O(D^{{1-p \over p}}), \ee where the inequality in the last line
holds for $L \le \m O(D^{1/p})$. Here $\mu$ is an arbitrary positive
constant, $L=8\pi\m N^2/\mu$, and \be D={4\kappa^{p-1}\eta\m N(17\m
N +4\mu) \over \m N(17-\eta) +4\mu(1-\eta)} \ee is obtained by
minimizing the denominator over $\bt$.

To get a tighter bound on $\m F^{-1}(0)$ we should take $\mu$ as
small as possible under the condition $L \le \m O(D^{1/p})$. For the
lossless case $\eta=1$, $\mu = \Theta(\m N^{\frac{2p}{p+1}})$. For
the lossy case $\eta <1$, $\mu = \Theta(\m N^{2-\frac{1}{p}})$.
These values of $\mu$ lead to \be \m F^{-1}(0) \ge {\bigg\{} \bx \m
O\lx (\m N/\kappa)^{2(1-p)\over p+1}\rx, \quad\ \rm{for} \quad\
\eta=1, \\ \m O\lx (\m N/\kappa)^{1-p\over p}\rx, \quad\ \rm{for}
\quad\ \eta<1 \ex \ee in the large $\m N$ limit.

Next we use a particular example to demonstrate the above results in
details. In the squeezing vacuum model in Ref. \cite{scien}, a phase
fluctuation is modeled by the spectrum $\tilde {\m
F}_C(\og)=\kappa/(\ld^2+\og^2)$, which is asymptotically identical
to the Wiener phase spectrum ($p=2$). For this beam, the Fourier
transform of $\Sigma(t-t')=4\me {\Delta n(t)\Delta n(t')}$ yields
\be \tilde \Sigma(\og) &=& 4\m
N+\frac{\ga^3}{16}\bigg[\frac{(R_+-1)^2(1-x)^3}{(1-x)^2\ga^2+\og^2}
\nn && + \frac{(R_--1)^2(1+x)^3}{(1+x)^2\ga^2+\og^2} \bigg], \ee
where the total photon flux is given by \be \m
N=\frac{\ga}{16}\lz(R_+-1)(1-x)+(R_--1)(1+x)\rz. \label{gamm} \ee
For an optical parameter oscillator, $R_\pm$ (for a squeezed vacuum
$R_+R_-=1$) are the anti-squeezing and squeezing levels at the
center frequency, respectively. Here $\ga$ is the cavity's decay
rate and $x=(\sqrt{R_+}-1)/(\sqrt{R_+}+1)$ is the normalized pump
amplitude.

To fulfil the numerical calculation, we take $\kappa=\ld=1$ and
$R_+=1/R_-=16 \m N^{1/3}$. The value of $\ga$ is solved by Eq.
(\ref{gamm}). Putting all equations together and performing the
integral, the final results after optimizing over $\bt$ are shown in
Fig. 3. It can be seen that the presence of photon losses always
blur the MSE and make a transitive behavior of the MSE from the
stochastic HL scaling $\m N^{-2/3}$ to the stochastic SQL scaling
$\m N^{-1/2}$ as the total photon flux increases.

\begin{figure}[t!]
\centering { \label{Fig.1}
\includegraphics[width=.9\columnwidth]{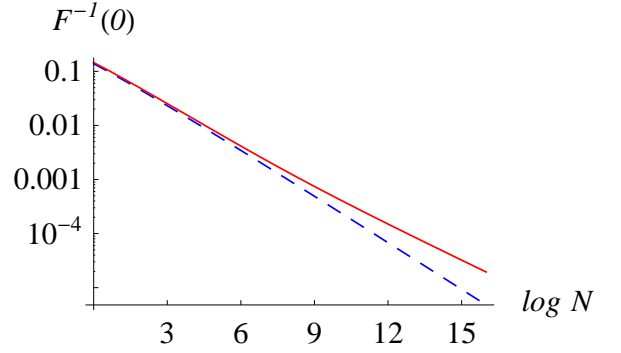}}\hspace{0.00in}
\caption{The plot of the upper bound to $\m F^{-1}(0)$ with respect
to the total photon flux $\m N$ for the squeezed vacuum model in
Ref. \cite{scien}. The solid (dashed) line denotes the result for
the lossy (lossless) case. Here $\eta=0.95$.}
\end{figure}

\section{Conclusion}

In summary, we have applied the variational method to obtain the
universal and analytical lower bounds for phase precision in some
noisy systems. We have derived a lower bound for phase precision in
lossy optical interferometry at non-zero temperature that allows us
to evaluate the effect of temperature on phase estimation. We have
also discussed the effect of both amplitude damping and phase
diffusion on phase-shift precision, which approaches to a constant
term even when the total available energy goes to infinity. At last,
we have extended the constant phase estimation to the case of
continuous fluctuating phase estimation, and have found that due to
photon losses the corresponding MSE transits from the stochastic HL
to the stochastic SQL as the total photon flux increases.

\begin{acknowledgments}
The author would like to acknowledge the support from NSFC Grand No.
11304265, the Education Department of Henan Province (No.
12B140013), and the Program for New Century Excellent Talents in
University (No. NCET-12-0698).
\end{acknowledgments}


\end{document}